# Don't Walk, Skip! Online Learning of Multi-scale Network Embeddings

Bryan Perozzi*, Vivek Kulkarni, Haochen Chen, and Steven Skiena
Department of Computer Science, Stony Brook University
{bperozzi, vvkulkarni, haocchen, skiena}@cs.stonybrook.edu

*Abstract*—We present WALKLETS, a novel approach for learning multiscale representations of vertices in a network. In contrast to previous works, these representations explicitly encode multi-scale vertex relationships in a way that is analytically derivable.

WALKLETS generates these multiscale relationships by subsampling short random walks on the vertices of a graph. By 'skipping' over steps in each random walk, our method generates a corpus of vertex pairs which are reachable via paths of a fixed length. This corpus can then be used to learn a series of latent representations, each of which captures successively higher order relationships from the adjacency matrix.

We demonstrate the efficacy of WALKLETS's latent representations on several multi-label network classification tasks for social networks such as BlogCatalog, DBLP, Flickr, and YouTube. Our results show that WALKLETS outperforms new methods based on neural matrix factorization. Specifically, we outperform DeepWalk by up to 10% and LINE by 58% Micro-F1 on challenging multi-label classification tasks. Finally, WALKLETS is an online algorithm, and can easily scale to graphs with millions of vertices and edges.

## I. INTRODUCTION

Social networks are inherently hierarchical. For example, in a human social network, each individual is a member of several communities, which range from small (e.g. families, friends), to medium (e.g. schools, businesses), to large (e.g. nation states, or all who share a common tongue).

As the scale of these relationships changes, so too can their topology. For example, consider a university student on a social network (as illustrated in Figure 1a). They may be very tightly tied to their friends and immediate family and form dense graph structures with these individuals (e.g. near cliques). However they will be more loosely tied to the average student at their university – in fact, they will have no direct social connection to the majority of their fellow students. Finally, they will have relatively few ties to all individuals in their nation, but still will share many attributes due to a common culture. Scale plays an important role in prediction tasks as well since they also vary from the specific (e.g. a user's movie interests) to attributes associated with the more general communities a user is a member of (e.g. past employers or schools).

Most prior works on network representation learning have treated this with a 'one-size fits all' approach, where a single network representation is used for all predictions made for a single user, thus failing to explicitly capture the multiple scales of relationships the node has within the network. In our view, it is desirable to have a family of representations which capture the full range of an individual's community membership.

\* Now at Google Research

In this paper, we study the problem of embedding the nodes within a large graph into a finite number of dimensions $d \in \mathcal{R}^d$ that capture the latent hierarchy of communities which a node participates in. These latent *multiscale representations* are useful for predictive learning tasks in social networks. At prediction time, the representations can be leveraged (both individually, or combined) to provide a more comprehensive model of the user's affiliations. The difference between similarity (distance in the latent space) over varying representation scales is illustrated in Figures 1b and 1c.

Recently, there has been a surge of interest in representation learning for social networks, primarily based around neural matrix factorization [1]–[3]. These methods have demonstrated strong task performance on semi-supervised network classification problems.

Despite their strong task performance, we find that these methods leave much to be desired. First, initial work only addressed the problem of learning representations at *multiple scales* indirectly, either as an artifact of the learning process [1] or via an unintuitive combination of different objective functions [2]. More recently, an approach for learning a multiscale representation [3] has been proposed, but its computational complexity renders it intractable for most real-world graphs. Second, most of these methods hinge on a 'one size fits all' approach to network representation learning, which masks the nuanced information present at each individual scale of the graph.

We propose WALKLETS, an online algorithm for learning social representations which captures multiple scales of relationships between vertices in a graph. Unlike existing work, WALKLETS's dimensions have meaning, allows for informed network classification and visualization of relationships captured. The method itself is scalable, and can be run on graphs with millions of vertices.

Specifically, our contributions are the following:
1) **Multiscale Representation Learning**: We propose a graph embedding algorithm which explicitly captures multiple scales of relationships.
2) **Evaluation**: We extensively evaluate our representations on multi-label classification tasks on several social networks (like BlogCatalog, DBLP, Flickr, YouTube, and Arxiv). WALKLETS outperforms several challenging baselines like DeepWalk [1], by up to 5 points of Micro-F1.

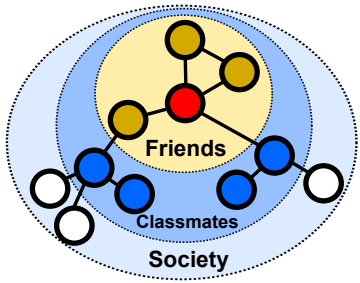 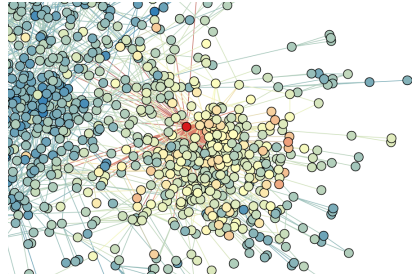 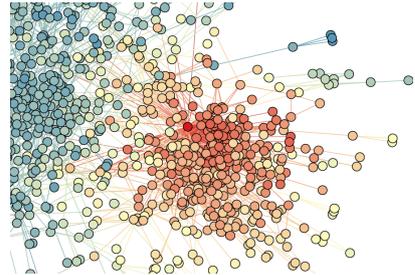

(a) A student (in red) is a member of several increasing larger social communities.

(b) WALKLETS Fine Representation

(c) WALKLETS Coarse Representation

Figure 1: Our method captures multiple scales of social relationships like those shown in Figure 1a. This is illustrated as a heatmap on the original graph in Figures 1b,1c. Color depicts cosine distance to a single vertex, with red indicating close vertices (low distance), and blue far vertices (high distance). Figure 1b: Only immediate are near the input vertex in a fine-grained representation. Figure 1c: In a coarse representation, all vertices in the local cluster of the graph are close to the input vertex. Subgraph from the `Cora` citation network.

3) **Visualization and Analysis**: WALKLETS preserves multiple scales of latent representations, which we use to analyzing the multiscale effects present in large graphs.

The rest of the paper is arranged as follows. In Section II-A we present a brief overview of representation learning for social networks, and expound on their use for graph classification problems. We follow this by analytically deriving the basis WALKLETS, our approach for Multiscale Social Representation Learning in Section III. We outline our experiments in Section IV, and present their results in Section V. We close with a discussion of related work in Section VII, and our conclusions.

## II. PRELIMINARIES

In this section, we briefly discuss necessary preliminaries of neural representation learning as it relates to our work.

### A. Notations and Definitions

We denote a network $G = (V, E)$, where $V$ represents the members of the network and $E$ their connections, $E \subseteq (V \times V)$. We refer to $G$'s adjacency matrix as $A$ where $A \in \mathbb{R}^{|V| \times |V|}$. The entry $A_{ij}$ is non-zero if and only if there is an edge $(i, j) \in E$. Given a graph $G$ with an associated adjacency matrix $A$, we define its view at scale $k$ to be the graph defined by $A^k$.

We define a representation as a mapping $\Phi\colon v \in V \mapsto \mathbb{R}^{|V| \times d}$ which maps every vertex to a $d$-dimensional vector in the real space. In practice, we represent $\Phi$ by $X$, a $|V| \times d$ matrix which is typically estimated from data. Finally, Let $G_L = (G, X, Y)$ denote a partially labeled social network, with features $X \in \mathbb{R}^{|V| \times d}$ where $d$ is the size of the feature space, and $Y \in \mathbb{R}^{|V| \times |\mathcal{Y}|}$, where $\mathcal{Y}$ is the set of labels.

### B. Problem Definition

**Multi-Scale Representation Learning**: Given a $G = (V, E)$, learn a family of $k$ successively coarser social representations, $X_1, X_2, \ldots, X_k$, where $X_k \in \mathbb{R}^{|V| \times d}$ captures the view of the network at scale $k$. Intuitively, each member of the family encodes a different view of social similarity, corresponding to shared membership in latent communities at differing scales.

**Predictive learning task**: Given a representation or a family of representations $\mathbf{X}$, learn a hypothesis $H$ that maps $\mathbf{X}$ to the label set $\mathcal{Y}$. This allows generalization (i.e. to infer labels for vertices in G which do not have labels).

### C. Representation Learning

Recall that the goal of representation learning is to infer a mapping function $\Phi\colon v \in V \mapsto \mathbb{R}^{|V| \times d}$. This mapping $\Phi$ represents the latent social representation associated with each vertex $v$ in the graph. (In practice, we represent $\Phi$ by a $|V| \times d$ matrix of free parameters, which will serve later on as our $X$).

In [1], Perozzi et al. introduces the concept of modeling a vertex as a function of its node co-occurrences in a truncated random walk. These sequences capture the diffusion process around each vertex in the graph, and encode local community structure. Consequently, the goal is to estimate the likelihood of a vertex $v_i$ co-occurring with its local neighborhood:

$$\Pr\left(v_i \mid \left(\Phi(v_1), \Phi(v_2), \cdots, \Phi(v_{i-1})\right)\right) \qquad (1)$$

However, as the walk length grows, computing this conditional probability becomes infeasible.

To address this computational complexity, it was proposed to relax the problem, by ignoring the order of neighboring vertices, and instead of using the context to predict a missing vertex, it uses one vertex to predict its local structure. In terms of vertex representation modeling, this yields the optimization problem:

$$\underset{\Phi}{\text{minimize}} \quad -\log \Pr\left(\{v_{i-w}, \cdots, v_{i+w}\} \setminus v_i \mid \Phi(v_i)\right) \qquad (2)$$

Interestingly, this optimization goal is directly equivalent to learning a low-rank approximation of the adjacency matrix, $A$.

## D. Neural Matrix Factorization

A commonly proposed approach to model the probability of a node $v_i$ co-occurring with $v_j$ uses a softmax to map the pairwise similarity to a probability space,

$$Pr(v_i|v_j) = \frac{exp(\Phi(v_i) \cdot \Phi'(v_j))}{\sum_{j \in \mathcal{V}} exp(\Phi(v_i) \cdot \Phi'(v_j))} \quad (3)$$

Where $\Phi'(v_i)$ and $\Phi'(v_j)$ represent the "input" and "output" word embeddings for node $v_i$ and $v_j$ respectively [4]. Unfortunately, calculating the denominator in Equation (3) is computational expensive.

Alternatively, noise-contrastive estimation (NCE) [5] has been proposed as a relaxation of Equation (3) [3], [4]. NCE models the probability of a vertex co-occurrence pair $(v_i, v_j)$ appearing as:

$$Pr((v_i, v_j)) = \sigma(\Phi(v_i) \cdot \Phi'(v_j)) = \frac{1}{1 + e^{-\Phi(v_i) \cdot \Phi'(v_j)}} \quad (4)$$

It can be shown that both of these probabilistic models correspond to implicitly factoring a transformation of the adjacency matrix [6]–[8].

*1) Social Representations as Matrix Factorization:* To motivate our method, we briefly discuss the matrices which prior work on learning social representations are implicitly factoring. Specifically, we discuss our closest related work - DeepWalk [1].

**DeepWalk**: A closed form solution which describes the implicit matrix factorization performed by DeepWalk is possible, as noted by [3], [8]. They derive a formulation for DeepWalk with Hierarchical Softmax (DWHS) showing that it is factoring a matrix $M$ containing the random walk transition matrix raised to the $t$ power. In other words, the entry $M_{ij}$ is the expectation of a path of length $t$ started at node $i$, ending at node $j$.

$$M_{ij} = \log \frac{\#(v_i, v_j)}{\#(v_i)} = \log \frac{[e_i(A + A^2 + \cdots + A^t)]_j}{t} \quad (5)$$

where $\#(v_i, v_j)$ is counting function, that counts the occurrences of pair $(v_i, v_j)$, and $e_i$ is a $|V|$-dimensional vector which serves as an indicator function (having a 1 in the $i$-th row, and 0s elsewhere).

## III. MULTI-SCALE NEURAL MATRIX DECOMPOSITION

In this section, we introduce our algorithm for creating multiscale network representations. We begin by describing our model for capturing different representations scales. We follow this with a discussion of concerns, like search strategy and optimization. Finally, we perform a case study on a small real-world citation network, which illustrates the different scales of network representations captured through our method.

### A. Model Description

Here, we build on the intuition developed in Section II-D1 and formally extend previous methods in social representation learning to explicitly model the multi-scale effects exhibited in real world networks.

*1) Multi-scale Properties of DeepWalk:* As shown in Equation (5), DeepWalk is implicitly factoring a matrix containing entries of $A, A^2 \ldots, A^k$, where $k$ is the window size over the random walk. Each power of $A^k$ represents a different scale of network structure (recall, the entries of $A_{ij}^k$ is the number of paths between nodes $i$ and $j$ of length $k$).

It is interesting to see then, that DeepWalk is already implicitly modeling dependencies of multiple scales. This shows that the representations learned are capable of capturing both short and long-distance dependencies between nodes in a social network. Although DeepWalk is expressive enough to capture these representations, its multiscale scale properties have limitations:

**Multiscale Not Guaranteed**: Multi-scale representations are not necessarily captured by DeepWalk, since they are not explicitly preserved by the objective function. In fact, since DeepWalk will always have more entries from A than from $A^k$, $(k > 1)$, it is *biased* towards representations that preserve the lowest power of $A$. To see this, observe that given any random walk of length $L$, the number of entries from $A$ is at most $L - 1$. In general, the number of entries from $A^k$ in a random walk is at most $\frac{L-1}{k}$.

This bias towards lower powers of the adjacency matrix is a fundamental weakness, when higher order powers are the appropriate representations for machine learning tasks on the network. For example, when classifying a feature correlated with large scale graph structure – such as language detection on a social network, a coarse-grained representation may offer performance benefits over a finer representation that preserves individual edges.

**Global Representation Only**: DeepWalk learns one global representation that conflates all possible scales of network relationships. As such, different scales of representation are not independently accessible. This is undesirable, as each individual learning task needs to decode the relevant similarity information from the global representation. In actuality, an ideal representation for social relationships would span multiple scales, each one capturing successively broader levels of latent community memberships. Each learning task then, is able to utilize the best level of social relationships for its task. As we will show in Section V, performance on each learning task can be maximized through a different scale of social representation.

### B. WALKLETS: *Multiple Scales of Random Walks*

Building on observations discussed so far, we propose extending the model introduced by [1] to explicitly model multiscale dependencies. Our method operates by factoring powers of the adjacency matrix $A$, using recently proposed algorithms for learning representations.

Similar to DeepWalk, we model the network through a series of truncated random walks started at each node. As discussed in [1], the co-occurrence of two vertices in these truncated random walks can model the rate of diffusion in the network. However, we make a key change to the sampling procedure. Specifically, we choose to *skip* some of the the nodes in the random walk. In this way, we form a set of relationships which

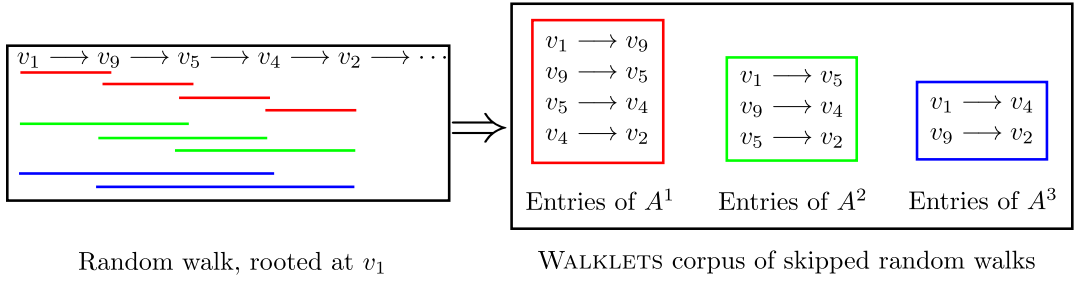

Figure 2: Overview of WALKLETS. Our method samples edges from higher powers of the adjacency matrix using a rooted random walk and skips over vertices. An edge sampled from $A^k$ represents a path of length $k$ in the original graph.

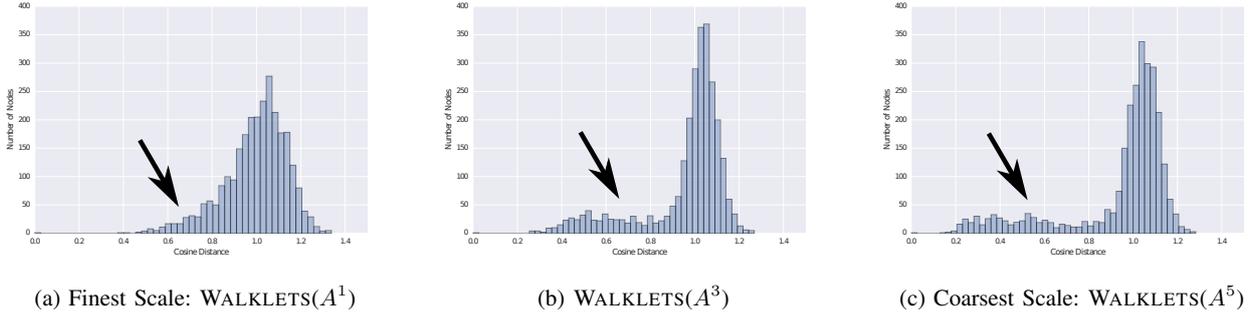

(a) Finest Scale: WALKLETS($A^1$)   (b) WALKLETS($A^3$)   (c) Coarsest Scale: WALKLETS($A^5$)

Figure 3: The distribution of distances to other vertices from $v_{35}$ in the Cora network at different scales of network representation. Coarser representations (such as WALKLETS($A^5$)) 'flatten' the distribution, making larger communities close to the source vertex. Graph heatmap of corresponding distances shown in Figure 4.

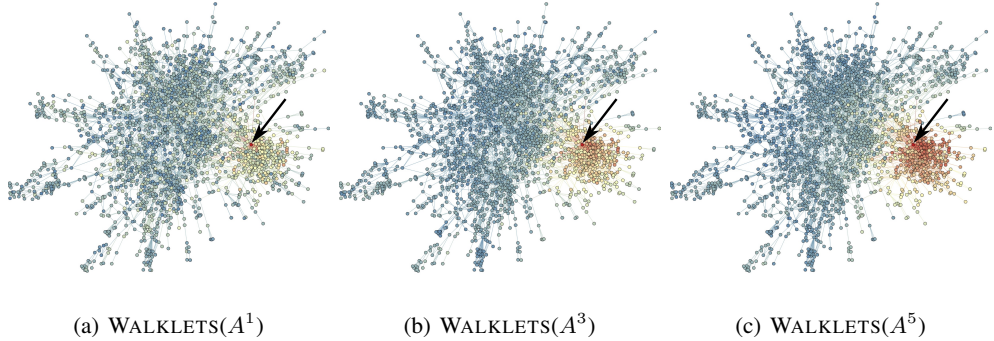

(a) WALKLETS($A^1$)   (b) WALKLETS($A^3$)   (c) WALKLETS($A^5$)

Figure 4: Heatmap of cosine distance from vertex $v_{35}$ (shown by arrow) in the Cora network through a series of successively coarser representations. Close vertices are colored red, distant vertices are colored blue. Corresponding distributions of vertex distances are shown above in Figure 3.

are sampled from successively higher powers of $A$. We denote WALKLETS($A^1, \ldots, A^k$) to represent WALKLETS derived from the $A^1, \ldots, A^k$ powers of the adjacency matrix respectively.

Each distinct power forms a corpus in a series of corpora which models a specific distance dependency in the network. This sampling procedure is illustrated in Figure 2.

After partitioning the relationships sampled by scale, we model the probability of observing a sample vertex $v_i$ with a vertex $v_j$ as in Equation 3. This implies the following objective function in order to learn representations for each node $v_i \in \mathcal{V}$:

$$J = -\sum_{v_i, v_j \in C_k} \log Pr(v_i|v_j) \qquad (6)$$

where $C_k$ is the corpus of random walk pairs generated for representation scale $k$. $J$ seeks to maximize the log likelihood of $v_i$ co-occurring with context node $v_j$. This objective, commonly referred to as *Skip-gram*, was first proposed for language modeling in [4], and first extended to network representation learning in [1].

*1) Loss function Optimization:* We optimize the loss function stated in Equation 6 using standard back propagation with stochastic gradient descent. We use the default learning rate as 0.025 and set the size of the embedding $d$ to 128 unless stated otherwise. This can be easily extended to weighted graphs (by adapting the gradient proportional to the weights). Techniques like edge sampling [2] can also be adapted to our method.

*2) Implicit matrix factorization view:* By sampling in this way, we can show that we learn representations of different scales. Using DeepWalk on skipped random walks, where the skip factor is set to $k$ (we sample nodes that are a distance $k$ from each other) implicitly factors a matrix derived from $A^k$.

Observe that in a skipped random walk with a skip factor of $k$, each consecutive node pair $v_i$ and $v_{i+1}$ are reachable by path of length exactly $k$ and therefore represent edges sampled from $A^k$. When we provide DeepWalk with walks where $v_i$ co-occurs with $v_j$ only if it is reachable by path of length $k$, in Equation 5, only the term corresponding to $A^k$ is present. Therefore we implicitly factor a matrix derived from $A^k$.

*3) Search Strategy:* Related work [1], [2], [9] have respectively advocated differing strategies for generating social relationships from edges in a graph. The *breadth-first* strategy successively expands from a single node of interest and examines all its neighbors. This works well for local neighbors, but faces a state space explosion as higher levels of expansion (i.e. neighbors of neighbors, etc). The *depth-first* strategy uses a random walk, which encodes longer distance information, and may be better for learning higher order network representations.

We have presented our method by observing relationships of multiple scales through random walk sampling, which we believe would be more scalable. It is worth mentioning that since node2vec [9] introduces a biased random walking scheme which could be seen as an extension to DeepWalk, our skipping algorithm could also be applied to the random walks it generates. An alternative search strategy (possible for smaller graphs), is to directly compute $A^k$ (i.e. all nodes with a path of length $k$ to another) and use this to sample pairs of vertices.

## C. Case study: `Cora`

In order to illustrate the effects of network representations at multiple scales, we visualize a small citation graph `Cora`, with 2,708 nodes and 5,429 edges.

Figure 3 shows a histogram of the distance from a particular node ($v_{35}$) to every other node's social representation. As we examine successively deeper social representations, we see that a group of *proximal* nodes develops. These nodes are part of the larger community of papers which $v_{35}$ is a member of – specifically, the area of `Genetic Algorithms`. Should we perform classification in `Cora`, this clear separation of network structure enables easier generalization.

This phenomenon is also illustrated by Figure 4, which shows a heatmap of distance to node $v_{35}$ overlaid on the original network structure. Note how the distance at different scales of network representation encodes membership in successively larger communities.

## IV. Experimental Design

In this section, we analyze our method experimentally by applying our method to several online social networks. In particular, our experiments are motivated by two main goals: First, we seek to characterize the various multi-scale effects manifested in different real world social networks. Second, we evaluate the efficacy of our method in capturing their underlying network structure.

We briefly provide an overview of the different graphs we use in our experiments in below:

- `BlogCatalog` is a network consisting of relationships between bloggers. The labels indicate the topic categories associated with authors. It contains 10,312 nodes, 333,983 edges, and 39 labels.
- `DBLP` is a co-author graph between researchers in computer science. The labels indicate the research areas that the researchers publish in. It has 29,199 nodes, 133,664 edges, and 4 labels.
- `Flickr` is a network consisting of users on the photograph sharing website `Flickr`. An edge in the network indicates a contact relationship between the user pair. The labels indicate the interest groups of the users (e.g. *noir photography*). It has 80,513 nodes, 5,899,882 edges and 195 labels.
- `YouTube` is a social graph between video enthusiasts. The labels indicate shared group memberships which users have in common (e.g. Anime videos). It has 1,138,499 nodes, 2,990,443 edges and 47 labels.

Since our method learns latent representations of nodes in a social network in an unsupervised manner, these representations should generally serve as useful features for a variety of learning tasks. Therefore, we evaluate our method on one such important task – that of multi-label classification in social networks. This task is motivated by observing that nodes in a network exhibit memberships in many of the same groups as their friends. For example, people in a social network are members of several circles (family, alma-mater, employer, shared hobbies, etc). Modeling and predicting these varied group memberships is not only essential to understanding real world networks, but also has several important commercial applications (e.g. more effective ad targeting).

### A. Baseline Methods

To compare with our approach, we consider three recently proposed models for social representation learning that represent the state of the art.

- **DeepWalk** [1]: This method learns representations of nodes using a sequence of truncated random walks. The learned representations capture a linear combination of community membership at multiple scales.
- **LINE** [2]: Similar to DeepWalk, this method learns node representations with the Skip-gram objective function. For this baseline we use the LINE 1$^{\text{st}}$ method, which only considers immediate connections (those in $A^1$).

Table I: Multilabel classification results on BlogCatalog and DBLP. Micro $F_1$ scores are reported. In general, specific higher order representations improve task performance. On DBLP, (a small, well behaved graph), the exact multiscale computation performed by GraRep outperforms WALKLETS's sampling approach. Numbers greater than DeepWalk are bolded. (*,**) indicates statistically superior performance to DeepWalk at level of (0.05, 0.001) using a standard paired t-test.

| % Labeled Nodes | 10% | 50% | 90% |
|---|---|---|---|
| WALKLETS($A^2$) | **37.46**** | **41.19*** | **42.59** |
| DeepWalk | 34.55 | 40.77 | 42.34 |
| LINE | 23.65 | 34.67 | 37.44 |
| GraRep | **37.05**** | **40.95** | 42.31 |
| Gain over DeepWalk(%) | 8.9 | 1.0 | 0.6 |
| Gain over LINE(%) | 58.4 | 18.8 | 13.8 |

(a) BlogCatalog

| % Labeled Nodes | 1% | 5% | 9% |
|---|---|---|---|
| WALKLETS($A^3$) | **60.49**** | **65.36**** | **66.24**** |
| DeepWalk | 54.93 | 63.45 | 65.16 |
| LINE | 45.03 | 51.69 | 53.32 |
| GraRep | **63.62**** | **67.47**** | **68.34**** |
| Gain over DeepWalk(%) | 10.1 | 3.0 | 1.7 |
| Gain over LINE(%) | 34.3 | 22.6 | 20.5 |

(b) DBLP

- **GraRep** [3]: This multi-scale method generates vertex representations by explicitly computing successive powers of the random walk transition matrix, and uses the SVD to reduce their dimensionality. GraRep is a very strong baseline. It is an exact computation similar to our approach in spirit, however, GraRep is not an online algorithm and does not scale to large graphs.

### B. Multi-label classification

We evaluate our method using the same experimental procedure outlined in [1]. We randomly sample $T_f$ fraction of the labeled nodes and use them as training data with the rest being used as a test data set. This process is repeated 10 times, after which we report the mean MICRO-F1 scores. We do not include the result on the other evaluation metrics like accuracy and MACRO-F1 since they all follow the same trend. This enables us to compare our method with other relevant baselines easily.

In all cases, we learn a logistic regression model (based on the *one vs rest strategy*) classification. We use a default $L2$ regularization penalty of $C = 1$ and use the optimization algorithm implemented by Liblinear [10].

For WALKLETS, we use only the representations generated from walks in $\pi \in \{1, 2, 3\}$. The number of walks $N$ from each node in all cases was set to 1000 while the length of each such walk $L$ is set to 11. The dimension of embeddings $d$ was 128 in all cases (these settings are also the same for all baselines). In cases, where we use more than one representation from $\pi$, we concatenate all such features and use PCA to project them down to 128 dimensions.

With this methodology, we control for differences in sizes of training data, and hyper-parameters that determine the capacity of the representations. This allows for an interpretable comparison with other methods and baselines.

## V. EXPERIMENTAL RESULTS

In this section, we present our results of the multi-label classification task on the various datasets described in Section IV.

### A. Multi-label Classification

Tables I and II show the performance of our algorithm and its gain over two state-of-the-art network embedding methods, namely DeepWalk and LINE. The gain over GraRep is not presented, since it is not an online algorithm and does not scale to large graphs (like Flickr and Youtube).

**BlogCatalog**: Table Ia shows the multi-label classification results for BlogCatalog. We observe that WALKLETS using features from $A^2$ outperforms all baselines with respect to Micro-F1. When labeled data is sparse, (only 10% of the nodes labeled), the difference is statistically significant at the 0.001 level for Micro-F1 (8.9% improvement over DeepWalk and 58.4% over LINE). Statistical significance was established using a paired t-test over the 10 different runs.

**DBLP**: The results from experiments on DBLP are shown in Table Ib. We notice that the representations of $A^3$ provide a statistically significant improvement of in Micro-F1 over DeepWalk and LINE. With 1% labeled nodes, the gain of WALKLETS($A^3$) over DeepWalk and LINE are 10.1% and 34.3% respectively. These coarser representations offer better encoding of the subject areas which an author publishes in.

We note that on this task however, WALKLETS fails to outperform the multiscale representation learned by GraRep. We attribute this to the fact that the DBLP is quite well behaved. First, it exhibits a highly homophilous behavior, as co-authorship guarantees similar attributes (a shared publication between two authors must be in the same research area). Second, co-authorship edges in the graph indicate a high degree of similarity (it is harder to create spurious edges). When such conditions hold, GraRep's direct computation of the random walk transition matrix can yield high performance gains. However, we note that GraRep's technique requires materializing a dense matrix, which is inherently unscalable – as illustrated by our remaining datasets.

**Flickr**: Table IIa shows the evaluation on the Flickr dataset. On this dataset, we see that features derived from $A^2$ offer the best performance on the task for Micro-F1 significantly outperforming all baselines. Specifically, we observe that when only 1% of the data is labeled for training, WALKLETS outperforms DeepWalk by 4.1% and LINE by 29.6% in Micro-F1 score.

We note that the most competitive baseline, GraRep, fails to run on this dataset (which has only 80,513 vertices) as it

Table II: Multi-label classification results in `Flickr` and `Youtube`. Micro $F_1$ scores are reported. Higher order WALKLETS representations outperforms all scalable competitors on these graphs. The most competitive baseline (GraRep) is unable to run on graphs with millions of vertices. Numbers greater than DeepWalk are bolded. (*,**) indicates statistically superior performance to DeepWalk at level of (0.05,0.001) using a standard paired t-test.

| % Labeled Nodes | 1% | 5% | 9% |
|---|---|---|---|
| WALKLETS($A^2$) | **32.47**** | **37.41**** | **38.70**** |
| DeepWalk | 31.18 | 36.64 | 38.18 |
| LINE | 25.06 | 30.55 | 32.85 |
| GraRep | - | - | - |
| Gain over DeepWalk(%) | 4.1 | 2.1 | 1.4 |
| Gain over LINE(%) | 29.6 | 22.5 | 17.8 |

(a) `Flickr`

| % Labeled Nodes | 1% | 5% | 9% |
|---|---|---|---|
| WALKLETS($A^2, A^3$) | **37.19*** | **40.73**** | **42.14**** |
| DeepWalk | 36.17 | 39.68 | 41.49 |
| LINE | 33.21 | 36.94 | 39.19 |
| GraRep | - | - | - |
| Gain over DeepWalk(%) | 2.8 | 2.6 | 1.6 |
| Gain over LINE(%) | 12.0 | 10.2 | 7.5 |

(b) `YouTube`

runs out of memory.[1] Our method, instead, handles such large networks gracefully while yielding competitive results. We discuss this further in Section VI.

**Youtube**: Our results for `YouTube` are presented in Table IIb. Once again, our method outperforms all baseline methods. Interestingly, the combined joint representation WALKLETS($A^2$,$A^3$) offers the best performance, significantly outperforming DeepWalk and LINE at the p=0.05 level. YOUTUBE contains much more vertices than any of the other datasets we have considered, and it is not surprising that GraRep again runs out of memory. We note that the online setting of WALKLETS allows learning of representations for graphs with millions of vertices.

## VI. DISCUSSION

Here we discuss our results further. We start by addressing the effect of different representation scales on classification tasks, move to discussing the scalability of our approach, and finish with an analysis of our sampling procedure.

### A. Multi-scale Effects in Classification

The experimental results presented in Section V show that no single representation scale provides the best performance across tasks. Providing explicit representations addresses a limitation shared by other baseline methods which do not explicitly encode information at multiple scales.

In terms of Micro-F1, features from $A^2$ performed best on two graphs (`BlogCatalog`, `Flickr`) at all variations of training data. This indicates that paths of length 2 are the appropriate social dependencies to model shared user interest (topic categories and photography interests, respectively). The neighbor-of-neighbor information captured in $A^2$ is especially useful to handle a network which is missing information. One important case where this occurs is the *cold start* problem, where a node has just joined the network and has not had the opportunity to make all of its connections yet.

Interestingly, representations derived from $A^3$ and ($A^2, A^3$) offered the best task performance on `DBLP` and `YouTube` respectively. On these graphs, the classification performance monotonically increases with the scale of representation captured. We can surmise that the classification tasks in these graphs do exhibit hierarchical structure, and that the distinct higher-order representations created through our method allow exploitation of the correlation between a graph's hierarchy and a task's labels.

We emphasize here that WALKLETS explicitly models the multi-scale effects present in social networks, enabling a comprehensive analysis of scales that are most informative for a given learning task – something which is not possible using methods which learn global representations [1], [2], or which blend representations of different scales together [3].

### B. Scalability

WALKLETS is an online algorithm which operates on pairs of vertices sampled at different dependency levels from the graph. This online approach approximates the higher-order transition matrices using sampling, which allows scaling to large graphs with millions of vertices. This is in stark contrast to the closest related method, GraRep, which require the materialization of a dense matrix ($A^k$ and similar matrices rapidly loose their sparsity as $k$ grows, if the graph is connected and the edge distribution follows a power law). The explicit need to compute the successive powers of the transition matrix is further complicated by GraRep's dependence on SVD – a computation that does not necessarily scale well to large networks.

### C. Sampling Analysis

Here we analyze the effectiveness of our proposed random walk sampling procedure by comparing it to the explicit and exact computation made by GraRep. Given a graph $G$, we use GraRep to explicitly compute the matrix $M_{GR}$ it factorizes. We then use the random walks used to learn WALKLETS embeddings to estimate the matrix WALKLETS factorizes $M_W$. We estimate how close our approximation is to the exact computation $M_{GR}$ by $Err = abs(M_{GR} - M_W)$. Parameter settings correspond to those described in Section IV-B.

In Table III, we report the mean error for an edge $(i, j)$ we observe in $Err$ for multiple scales on the two graphs small enough to explicitly compute the random walk transition matrix: `DBLP` and `BlogCatalog`. We observe that the mean error and the standard deviation of our approximation at various scales is low, indicating that our method captures a reasonable approximation of the random walk transition matrix. Increasing

---
[1]Our machine used for experiments was generously apportioned with 384GB RAM.

Table III: Mean and Standard Deviation of errors observed from sampling the transition matrices using WALKLETS as compared to exact computation.

|       | BlogCatalog $Err_{ij}$ |                      | DBLP $Err_{ij}$     |                      |
|-------|------------------------|----------------------|---------------------|----------------------|
|       | Mean                   | Std.Dev              | Mean                | Std.Dev              |
| $A^1$ | $9.3 \times 10^{-5}$   | $3.70 \times 10^{-3}$ | $1.4 \times 10^{-5}$ | $1.69 \times 10^{-3}$ |
| $A^2$ | $1.3 \times 10^{-4}$   | $7.64 \times 10^{-4}$ | $1.8 \times 10^{-5}$ | $1.20 \times 10^{-3}$ |
| $A^3$ | $1.5 \times 10^{-4}$   | $4.48 \times 10^{-4}$ | $2.0 \times 10^{-5}$ | $1.06 \times 10^{-3}$ |

the sample size (by adding random walks) will result in increasingly better approximations.

## VII. RELATED WORK

The related work falls into three categories: unsupervised feature learning, multiscale graph clustering, and graph kernels.

**Unsupervised Feature Learning** Recently proposed methods for social representation learning [1], [2], [9], [11] use neural network losses proposed by [4] to learn representations which encode vertex similarity. Unlike our work, these methods do not explicitly preserve the multiscale effects which occur in networks. The closest related work to ours [3], explicitly models multiscale effects in networks through an SVD on the random walk transition matrix. Our work uses sampling and online learning to operate on much larger corpora. In addition, our work preserves multiple scales of representation, which can provide both better task performance and modeling insight.

Distributed representations have also been proposed to model structural relationships in diverse fields such as computer vision [12], speech recognition [13], and natural language processing [14], [15].

**Multiscale Community Detection** Many techniques for detecting multiple scales of discrete communities in graphs have been proposed [16]–[18]. In general, unlike our method, these approaches seek to return a hard clustering which describes the hierarchy of social relationships in a graph.

**Graph Kernels** Graph Kernels [19] have been proposed as a way to use relational data as part of the classification process, but are quite slow unless approximated [20]. Recent work has applied neural networks to learn subgraph similarity for graph kernels [21].

## VIII. CONCLUSION

In this work, we introduce multi-scale network representations to specifically capture network structure at multiple resolutions. The online algorithm we propose, WALKLETS, uses the offsets between vertices observed in a random walk to learn a series of latent representations, each of which captures successively larger relationships. WALKLETS utilizes connections between neural representation learning algorithms and matrix factorization to theoretically underpin the matrices implicitly factored at each scale. We demonstrate empirically that WALKLETS's latent representations encode various scales of social hierarchy, and improves multi-label classification results over previous methods on a variety of real-world graphs.

In addition to strong performance, WALKLETS scales gracefully to arbitrarily large networks.

We believe that WALKLETS's explicit handling of multiple scales of representations allows better comprehension of each network's nuances, and that it lays a strong foundation for developing future multi-scale network approaches. Our further investigations in this area will seek to develop additional theoretical backing for our methods.